\documentstyle[12pt]{article}
\makeatletter
\newcommand{\singlespacing}{\let\CS=\@currsize\renewcommand{\baselinestretch}
{1.0}\tiny\CS}
\newcommand{\doublespacing}{\let\CS=\@currsize\renewcommand{\baselinestretch}
{1.5}\tiny\CS}

\begin{document}
\title{The Study of Geometric Phase in Twisted Crystal}
\author{Dipti Banerjee \\
\singlespacing Department of Physics \\
Rishi Bankim Chandra College \\
Naihati,24-Parganas(N)\\
Pin-743165, West Bengal\\
INDIA}
\date{}
\maketitle

\centerline{Abstract}
 The polarization matrix ($2\times2$)
obtained from two component
 eigen-spinors of spherical harmonics help us to evaluate the
 differential matrix $N$ of the anisotropic optical medium.
 The geometric phase is realized through {\it helicity} of photon, assuming the
 transmission of polarized light through the crystal which has been twisted
  about the normal to its surface over a closed path.

\vspace{2cm} Email: deepbancu@hotmail.com
 \vspace{4cm}

 \pagebreak

  Optical anholonomy was discovered long before in 1950 by Pancharatnam[1]
  in his strikingly original work.  From classical point of view Pancharatnam
  obtained his phase -$\Omega©/2$ developed by the cyclic change of polarizat
  ion of light over a closed path on the Poincare sphere.Berry from quantum
  mechanical point of view studied the {\it Geometric Phase}(GP) of linearly
  polarized light representing photon as two component spinor [2]. Twisting
   the anisotropic medium over a closed path, Berry also [3]studied the
  phase two-form (GP) in connection with the dielectric tensor and
  bierfringence of the medium. Recently Berry and Klein [4], pointed out
  that the passage of polarized light through twisted stack polarizers
  ($P$) or retarders ($R$) produces the geometrical and"geometrical +
  dynamical phases respectively,"   as cycled around the Poincare sphere
  on $N$ similar arcs. The geometric phase in the context of polarization
  optics have been studied from the view-point of group theoretical
  aspects by Simon and Mukunda [5] and Bhandari [6].  In the light of
  Berry's work we have recently evaluated the polarization matrix $M$
  of an optical system from ortho-normal eigenvector represented by
  two-component spinors of spherical harmonics. We also obtained the
  nonzero geometric phase of polarized light whose plane of
  polarization  is rotated over a closed path by a rotator [7]. In the light of
  Jones [8], here  we will calculate from the polarization matrix $M$,
  the optical property of the medium at each point, represented by
  differential matrix $N$. The optical devices through which light
  passes are twisted such a way that the polarized light suffers
  variation over a closed path. This results to calculate the
  {\it Geometric Phase} of a single photon through the
   net change of inclination of helicity over a closed path
   in the relativistic framework.

  This present work is based on the consideration of light passing normally
  through a crystal that has been twisted about the axis parallel to the
   direction of transmission of the light. By means of an optical system
  of polarization matrix $M$ [8], we will study the properties of
  differential matrix $N$ which refer not to the complete element,but
  only to a given infinitesimal path length within the element.
  In fact the matrix operator $M$ of the complete element is
  represented as a line integral of a matrix $N$, which describe the
  optical properties (dielectric and gyration tensors) of the medium.
 The transmission of light through optical device [9] results the change
  of  polarization state which mathematically can be represented as
\begin{equation}
 \epsilon=M\epsilon_o
\end{equation}
If the property of light with the passage through any optical
element remain unaltered then the state is the eigenvector of the
optical component and in the language of matrix,
 Jones had shown the condition
\begin{equation}
M^n \varepsilon_i=d_i\varepsilon_i
\end{equation}
 where $d_i$ is the constant known as eigenvalue corresponding to
  the eigenvectors $\varepsilon_i$. For one component optical element the matrix
  becomes $ M=\pmatrix{m_1&m_4\cr m_3&m_2}.$\\
 For a given direction and for a given wavelength the vibration at the plane $z+dz$ is given by
\begin{equation}
\frac{d\epsilon}{dz}=\frac{dM}{dz}\epsilon_0
=\frac{dM}{dz}M^{-1}\epsilon
\end{equation}
where it is evident that $N$ is the operator that determines
$dM/dz$ from $M$ as follows
\begin{equation}
N=\frac{dM}{dz}M^{-1}
\end{equation}
This implies that the polarization matrix is the integral of
differential matrix $N$
\begin{equation}
M=M_0\exp(\int{Ndz})
\end{equation}

 When originally homogeneous crystal is twisted uniformly about an axis
  parallel to the direction of transmission of light, the differential
   matrix shows its special dependence upon $z$.
  Jones [10] had pointed out that the $N$-matrices are transformed upon rotation
 of the element just as the same way as the polarization matrix $M$ as bellows
 \begin{equation}
 N=S(kz)N_0 S(-kz)
  \end{equation}
 where $S$ is the rotation matrix and $k$ is the angular twist per unit
 thickness and if $N_0$ be the matrix corresponding to the untwisted crystal.
  Then the $z$ independent differential matrix $N'$ becomes
 \begin{equation}
  N^\prime=N_0 - kS(\pi/2)
 \end{equation}

This matrix of the twisted element satisfy the equation relating
the twisted state $\epsilon'$ as follows
\begin{equation}
\frac{d\epsilon'}{dz}=N'\epsilon'
\end{equation}

 The representation of photon as spinor had been observed in the earlier work of Tomita
 and Chaio in 1986 which further extended and supported by Berry himself.
Photons have no magnetic moment and so can not be turned with a
magnetic field. But they have the helicity along their propagation
direction which plays the important role   in our work. Here the
polarized photon has been represented relativistically by two
component spinor of spherical harmonics where the effect of
helicity is visualized by the parameter $\chi$ [11]. In fact we
have suggested that a photon with a fixed helicity can be viewed
as if a direction vector $y_\mu$ is attached at the space-time
point $x_\mu$ in Minkowski space, so that we can write the
coordinate in the complexified space-time as $z_\mu=x_\mu+ i
y_\mu$. Thus the wave
function$\phi(z_\mu)=\phi(x_\mu)+i\phi(y_\mu)$ should take into
account the polar coordinates $r,\theta,\phi$ along with the angle
$\chi$ which specify the rotational orientation around direction
vector $y_\mu$. Apart from the eigenvalue $m$ and $l$ of the
respective parameter $\theta$ and $\phi$, there exists $\mu$  for
$\chi$ which is the measure of anisotropy, and is given by the
eigen value of the operator $i \frac{d}{d\chi}$ having eigenvalues
$\mu=0,\pm1/2, \pm1....$ .In a three-dimensional anisotropic
space, it has been constructed following Fiertz[12] and Hurtz[13]
the spherical harmonics for $(l=1/2, m =\pm1/2 $ and $\mu=\pm1/2
$) half orbital angular momentum. Further it has been shown that
the two component spinor and its conjugate state can be formed
from the above spherical harmonics. In this paper we have
represented relativistically the polarized photon by two component
spinor of spherical harmonics where the effect of {\it helicity}
is visualized by the parameter $\chi$.  The behavior of chiral
photon with a fixed {\it helicity} $\pm 1$ in the polarized light
is similar to massless fermion having {\it helicity} +1/2 or -1/2.

In the light of Jones, the following polarization matrix $M$
 with these eigen-spinor of spherical harmonics
\begin{equation}
\begin{array}{lcl}
M=TDT^{-1}=1/2\pmatrix{-\cos\theta & \sin\theta e^{-i\chi}\cr
\sin\theta e^{i\chi} & \cos\theta \cr}
\end{array}
\end{equation}
represented by each point on the Poincare sphere parameterized by
the angle $\theta$ and $\chi$. Here $T$ is the matrix formed by
the orthonormal eigenvecters of spherical harmonics and $D$
denotes the eigenvalue matrix that reflects here the helicity
(+1/2,-1/2) of the polarized photon.

Our intention is to find out the geometric phase (GP) from the
known differential matrix $N$.
 This GP is concerned with the twisting of the optical medium about an axis along
 the direction of propagation of the incident light.
At a particular position of $z$, the $N$ matrix is related with
$M$ as follows
\begin{equation}
N=(\frac{dM}{d\theta})(\frac{d\theta}{dz})M^{-1}
\end{equation}
where $z=cos\theta$ Substituting the matrix value of $M$ (equation
28) in the above equation we find
\begin{equation}
N=1/{2\sin\theta} \pmatrix{0 & e^{-i\chi} \cr -e^{i\chi} & 0}
\end{equation}

The above differential matrix will resembles to the matrix of the
birefringent plate when the initial position of the system is
chosen as $z=0$ or $\theta=\pi/2$. Thus we have
\begin{equation}
N_0=1/2 \pmatrix {0 & e^{-i\chi} \cr -e^{i\chi} & 0} =(i/2)
\sigma_2 \pmatrix {e^{i\chi} & 0 \cr 0 & e^{-i\chi}}
\end{equation}
where $\sigma_2$ is the Pauli matrix and the other matrix is
representing a compensator
 whose slow axis is along the transverse $x_1$ direction such that
it introduces a phase difference $\chi$ between $E_1$ and $E_2$
components of the (transverse)
 electric field vector. The corresponding eigenvectors of the differential matrix $N$ are

\begin{equation}
|{\vec D}>=1/2 \sin\theta {\pm i \choose -e^{i\chi}}
\end{equation}
having the eigenvalues
\begin{equation}
\lambda=\pm i/2 \sin\theta
\end{equation}
that represent the polarised light on the Poincare sphere
parametrised by the angle $\theta$ and $\chi$. A polarization
state is cycled over a closed loop C on the Poincare sphere. The
circuit will be accompanied by a geometric phase $\gamma$,
\begin{equation}
\gamma = arg<\psi_{initial}|\psi_{final}>
\end{equation}
by twisting the medium through one complete turn along the path of
the beam. The geometric phase for the two {\it helicities}
$(\sigma.e_{k})=\pm1$ of the polarized photon would be
\begin{equation}
\gamma_{\pm1}©=\pm \Omega©
\end{equation}
where $\gamma$ is the solid angle swept out by $e_{k}$ on the
sphere.

Using Jones idea [14] we assume the twist of the optical element
at a particular position along the direction of the beam, so that
it causes a shift of the differential matrix from the initial
value  $N_0$ to $N^{\prime}$ given by
\begin{equation}
N^\prime=1/2 \pmatrix{0 & e^{-i\chi} \cr -e^{i\chi} & 0} - k
\pmatrix{0 & -1 \cr 1 & 0} = \pmatrix{0 & (1/2) e^{-i\chi} + k \cr
-(1/2)e^{i\chi}-k &0}
\end{equation}
where $k$ is the angular twist per unit thickness. This leads the
initial eigen
 state depart from the final. Using eqn (13) and the initial eigenstate
  as
 $$|\vec{D_0}>=1/2{i \choose -e^{i\chi}}$$
the twisted final state becomes
\begin{equation}
|\vec{D^\prime}>= ½ {i-\cos\theta(k e^{i\chi}+1/2) \choose ·
i\cos\theta(k+1/2e^{i\chi})-e^{i\chi}}
\end{equation}

 The interference between the initial and the final state
 originates the required geometric phase.
\begin{equation}
<\vec{D_0}|\vec{D^\prime}>=1/4[2+i\cos\theta(2k\cos\chi+1)]
\end{equation}
This shows that the required geometric phase is represented by
$\cos\chi$ when the parameter is twisted over a closed path. It
can be noted that this variation of angle $\chi$ is associated
with the change of direction of helicity of the photon, when the
optical device is twisted at a point along the direction of
propagation of light. Our early discussions suggest that $\mu$ as
well as $\chi$ is associated with a change of angular momentum
[7]. Hence, it means that the polarization of the light can be
changed with the change of angle $\chi$ which is associated with a
transfer of angular momentum between the optical system and the
incident light. This idea supports the analysis given by Tiwari
[15]. In view of this, the total phase change is obtained by the
parallel transport of the angle $\chi$ (angle of inclination of
{\it helicity})  over a closed path that indicates a rotation of
the plane of polarization.
\begin{equation}
\gamma=\int^{2\pi}_0 ¼ [2 + i\cos\theta (2k\cos\chi +1)]d\chi =
\pi=1/2(2\pi)
\end{equation}
Comparing with the phase factor $\gamma=(1/2)\Omega$, of Berry and
Pancharatnam [1,2],we have found here that the solid angle
associated with the GP is $2\pi$. This explains again that a
complete rotation of spin-$1/2$ matrix is accomplished by rotating
$\chi$ over a closed path $0 \le \chi \le 2\pi$ corresponds to
only a half turn of the photon,which is the reverse of the field.

I like to conclude that in the light of Jones calculus, the
differential matrix $N$ which is the optical property of the
complete element (matrix M) at an point along the infinitesimal
path  length  of the propagation has been determined in terms of
the variables $\theta$ and $\chi$ where the latter defines the
inclination of {\it helicity} of the polarized photon. Our
required geometric phase is obtained as the optical device is
twisted such a way at a particular point along the direction of
the propagation that the polarized light  varies over a closed
path with the simultaneous rotation of {\it helicity} of the
chiral  photon. This is the very source of relativistic geometric
phase  of the two state quantum photon.

{\bf Acknowledgement} I express my gratitude to all the authors in
my reference.

\vspace{1.00cm}
\section{\bf Reference}
\singlespacing
\begin{enumerate}
\item [1] S.Pancharatnam; Proc.Ind.Acd.Sci:{\bf A44},247,(1956).
\item [2] M.V.Berry; J.Mod.Opt:{\bf 34},1401,(1987).
\item  [3] M.V.Berry; Proceedings of a NATO Advanced Research Workshop
on Fundamental Aspects of Quantum Theory, held in Italy, Sep.
1985; Lectures presented at the International School on
"Anomalies, Phases, Defects...held in Ferrara, Italy, June 1989."
\item  [4] M.V.Berry and S.Klein; J.Mod. Opt.{\bf 43}, 165, (1996).
\item  [5] R.Simon and N.Mukunda; Phys.Lett{\bf 138A},474,(1989).
\item  [6] R.Bhandari; Phys. Lett{\bf 157},221,(1989).
\item  [7] D.Banerjee; Phys.Rev.-E{\bf 56}July, 1129 (1997).
\item  [8] R.C.Jones; J.Opt.Soc.Am.{\bf 31} 488 (1941).
\item  [9] W.A.Shurcliff;{\it Polarized Light and Use}
(Cambridge University,Havard University Press; 1962).
\item  [10] R.C.Jones; J.Opt.Soc.Am {\bf 38},671,(1948).
\item  [11] P.Bandyopadhyay; Int.J.of Mod.Phys: {\bf A4}, 4449,(1989);
{\it Geometry,Topology and Quantization}(Kluwer Academic
Publisher, 1996, The Netherlands).
\item  [12] M.Fierz; Helv. Phys. Acta: {\bf 17}, 27 (1944).
\item  [13] C.A.Hurst; Ann.Phys:{\bf 50}, 31, (1968).
\item  [14] R.C.Jones; J.Opt.Soc.Am,{\bf 46}, 126, (1956).
\item  [15] S.C.Tiwari; J.Mod.Opt: 39,1097, (1992).

\end{enumerate}

\end{document}